\documentclass[fleqn,10pt]{arXiv}
\pdfoutput=1

\usepackage[english]{babel}
\usepackage{natbib}
\usepackage{float}
\usepackage{textcomp}

\usepackage{pdfpages}  

\definecolor{color1}{RGB}{120,0,0}
\definecolor{color2}{RGB}{0,20,20}
\definecolor{color3}{RGB}{0,0,0}

\usepackage{hyperref}
\hypersetup{hidelinks,colorlinks,breaklinks=true,urlcolor=color1,citecolor=color1,
            linkcolor=color1,bookmarksopen=false,pdftitle={Title},pdfauthor={Author}}

\JournalInfo{Reply}
\Archive{}

\PaperTitle{Not even wrong: Reply to Wagg et al. \\
\vspace{0.25cm} \normalsize{Reply to ``Not even wrong: Comment'' by Wagg et al. (2019)}\thanks{This preprint represents
our official reply to the \citet{wagg.etal_19} Comment. A version of this reply was rejected for publication
in \textit{Ecology} for non-scientific reasons.}}
\PaperSubtitle{Reply to Wagg et al. Comment}

\Authors{Pradeep Pillai\textsuperscript{1,$\dagger$} and Tarik C. Gouhier\textsuperscript{1}}
\affiliation{\textsuperscript{1}\textit{Marine Science Center, Northeastern University, 430 Nahant Road, Nahant, MA 01908}}
\affiliation{\textsuperscript{$\dagger$}\textbf{Corresponding author}: pradeep.research@gmail.com}
\newcommand{\keywordname}{Keywords}

\Abstract{We demonstrate that the issues described in the \citet{wagg.etal_19} Comment on our paper \citep{pillai.gouhier_19} are all due to misunderstandings about the implications of pairwise effects, the nature of the null baseline in both our framework and in the Loreau-Hector (LH) partitioning scheme (i.e., the midpoint of the monocultures), and the impact of nonlinearity on the LH partitioning results. Specifically, we show that (i) pairwise effects can be computed over any time horizon and thus do not imply stable coexistence, (ii) the midpoint of the monocultures corresponds to a neutral community so coexistence was always part of the LH baseline, and (iii) contrary to what Wagg et al. suggested, generalized diversity-interaction models do not account for (and may in fact exacerbate) the problem of nonlinearity in monocultures, which inflates the LH net biodiversity effect and generates incorrect estimates of selection and complementarity. Hence, all of our original claims about the triviality inherent in biodiversity-ecosystem functioning research and the issues with the LH partitioning scheme hold.}

\begin{document}
\flushbottom
\maketitle
\thispagestyle{firstpage}


\section*{Introduction}

\citet{wagg.etal_19}'s Comment on our paper ``Not even wrong'' \citep{pillai.gouhier_19} misconstrued our critique of biodiversity-ecosystem functioning (BEF) research by insisting that it was rooted in the notion that all ecosystem effects were attributable to stable species coexistence. After erecting this strawman argument, Wagg et al. proceeded to knock it down by summarizing the history of the BEF research program and its many important distinctions from ``coexistence theory''. This was followed by two sections showing that stable coexistence can occur without overyielding, and vice versa. Unfortunately, the issues raised by Wagg et al. are all due to serious misunderstandings about the claims that were made in our paper. Furthermore, many of the statements in their Comment were justified by quoting sentence fragments from our paper that had been shorn of critical clauses and qualifiers such as ``default expectation'' or ``all things being equal''. This led to the scurrilous use of fabricated quotes that had us making statements like ``species in mixtures coexist and by some form of niche partitioning overyield'', when no such text appeared in that form anywhere in our original paper. Although the Wagg et al. Comment is largely rendered moot by the fact that we neither stated nor implied that all BEF effects could be ascribed to stable species coexistence, we nonetheless explicitly address all of its claims and criticisms below.

\section*{BEF and coexistence}
The first part of the Wagg et al. Comment consists of a thorough recounting of BEF's historical and agricultural origins that has no bearing on any of the philosophical or mathematical issues that were raised in our critique. Its entire purpose seems to have been to justify the false claim that we had ``[assumed] that BEF relationships \emph{must be positive} due to niche partitioning according to CT [coexistence theory]'', and thus our critique ``[overlooked] the rich history of research on the effects of species mixing and its purpose which is independent of CT'' (emphasis added). However, because we never made such a claim, this background information about the history of BEF was both unnecessary and irrelevant.

Additionally, Wagg et al.'s historical survey contained a number of confused remarks, as well as dubious and unsubstantiated claims about key concepts in BEF. For example, they stated that \citet{may_72}'s demonstration of the ``negative relationships between ecosystem complexity and stability'' in food webs was, in part, the source of the ``initial skepticism against positive BEF relationships''. However, it is hard to fathom how May's analytical results demonstrating the negative effects of increasing ecosystem complexity on the asymptotic or long-term stability of food webs apply to experimental assessments of the (according to Wagg et al.) transient or short-term relationship between species richness and ecosystem functioning within competitive guilds.

\begin{figure*}[!htb]\centering
\includegraphics[width= 0.7 \linewidth]{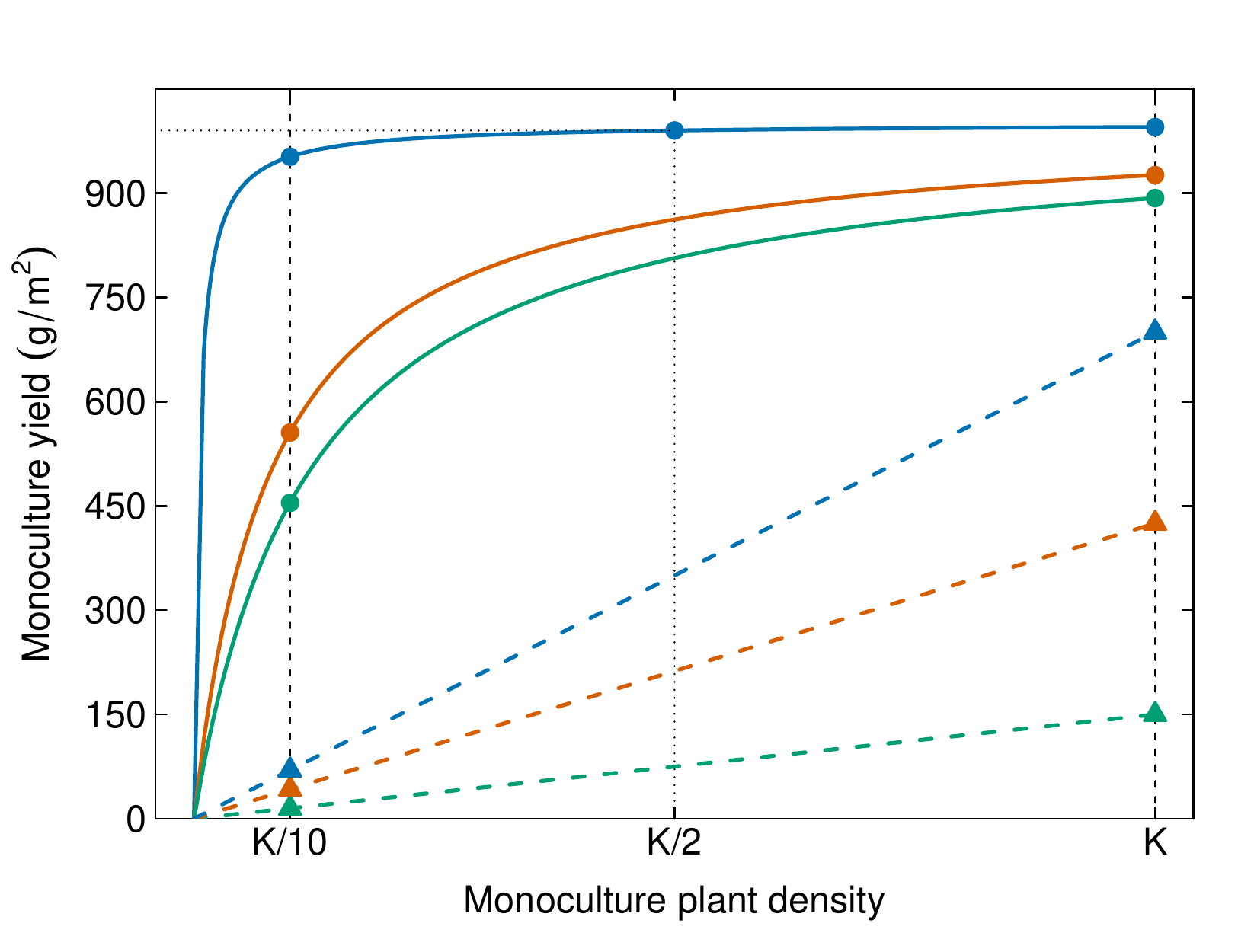}
\caption{\textbf{Dispelling misconceptions about the law of constant final yield}. (a) The effects of increasing monoculture density from 0 to carrying capacity $K$ on monoculture yield for three species represented via different colors. Under the law of constant final yield, all three species attain their maximum monoculture yield at relatively low densities due to the nonlinear (increasing and saturating functional) relationship between density and yield (solid curves). In the absence of this law, yield increases linearly with density in monocultures (dashed lines). The sampling effect does not require the law of constant final yield to hold, as differences in the yield of species (filled triangles and circles) at low ($K/10$) or high ($K$) sowing densities can potentially lead to elevated ecosystem functioning in mixtures.}
\label{fig1}
\end{figure*}

Similarly, Wagg et al. made several incorrect or unclear claims regarding the ‘law of constant final yield', which they stated was necessary for the sampling effect to emerge, or for relative yield totals (RYT) to be equal to 2 in the absence of facilitation. However, the sampling effect does not require the `law of constant final yield', which describes the increasing and saturating functional relationship between plant monoculture yield and sowing density. Indeed, as long as species differ in their monoculture yields, a sampling effect can potentially arise in mixtures whether the functional relationship between monoculture yield and sowing density is linear (\textit{contra} the `law of constant final yield') or nonlinear (Fig. \ref{fig1}).

Wagg et al. also stated that the ``concept of constant final yield in monocultures may be akin to the concept of a species carrying capacity ($K$) used in species competition-coexistence frameworks''. This statement is false: the `law of constant final yield' actually prevents such a facile analogy between monoculture yields and carrying capacities. Indeed, under the law of constant final yield, multiple plant sowing densities will map onto the same monoculture yield, thus making it impossible to equate monoculture yields and carrying capacities (Fig. \ref{fig1}: e.g., $K/2$ and $K$ map onto the same monoculture yield for the solid blue curve). This misunderstanding could help explain the labeling problems in Wagg et al.'s supplementary figure. For instance, in panel (h) depicting non-interacting species in mixtures ($\alpha_{i,j} = 0$), the carrying capacities $K_i$ and $K_j$ are projected onto the ecosystem functioning axes (yield), thus implying that these fundamentally different quantities are equivalent and have the same units (the isoclines are also confusingly labeled $K_i/\alpha_{i,j}$, implying division by zero).

Overall, the law of constant final yield is neither necessary for the sampling effect to emerge, nor sufficient to suggest that carrying capacities and monoculture yields can be treated analogously. Ironically, Wagg et al. overlooked the real implications of the law of constant final yield for BEF, which are the inflation of the Loreau and Hector (LH) net biodiversity effect and the consequent spurious estimation of selection and complementarity \citep{pillai.gouhier_19}.

The next two sections of Wagg et al.'s Comment showed that coexistence can occur without overyielding, and that overyielding can occur without stable coexistence. Regarding the former, the fact that Wagg et al. thought it important to list the many ways that coexistence can lead to no overyielding suggests two things. First, Wagg et al. incorrectly believed that we were arguing that coexistence necessarily results in overyielding. Second, by listing the many departures from null expectations that can be observed in nature, Wagg et al. seem to have misunderstood the whole purpose of a baseline. The fact that mixtures of three or more species may produce yields that differ from our null expectation based on pairwise mixtures merely means that there is something unique and interesting about the degree of ecosystem functioning observed in species-rich communities that cannot be captured by extrapolating from the simplest (pairwise) mixtures. It is precisely these kinds of interesting phenomena that our adaptive baseline is designed to detect. This stands in stark contrast to the LH baseline (the midpoint of the monocultures), which cannot distinguish the trivial effects of coexistence predicted by more than a century of basic ecological theory, from the more interesting higher order effects.

Similarly, Wagg et al.'s point that overyielding can occur without stable coexistence is irrelevant, since nothing about stable coexistence is implied by our critique; coexistence was always understood as being defined within the time frame of the experiment being conducted. Our baseline expectation was that, on average, a set of species will likely exhibit some degree of niche non-overlap due to the nature of niches as deduced from ecological theory. Nowhere is stable (asymptotic) coexistence implied, assumed or required. More importantly, their point regarding ecosystem functioning and stable coexistence is not even particularly interesting.

Take for instance their reference to the \citet{turnbull.etal_13} study, which showed how transient plant coexistence can result in overyielding. Here some degree of fitness stabilization (niche differentiation) will allow a net biodiversity effect to arise, even when long-term coexistence is impossible. The same self-evident point could be demonstrated by using essentially any unstable two-species model (e.g., Lotka-Volterra) with sufficiently large initial abundances. Given some suitable mix of fitness stabilizing and equalizing effects, and a short enough time frame, one can always detect transient biodiversity effects on the way to competitive exclusion. If anything, increasing niche differences are only likely to widen the window in which such biodiversity effects can be detected. As such, these results are just artefacts of how arbitrarily varying the time frame of the experiment will give different estimates of biodiversity effects. In some ways, Wagg et al.'s emphasis on the ways in which transient coexistence can give rise to positive biodiversity effects undermines the BEF research program even more acutely than our critique. Indeed, their argument underscores just how easy it can be to obtain essentially any desired (usually positive) result so long as the time frame is sufficiently short, particularly when starting with an oversaturated community, as most BEF experiments do.

\section*{Dispelling misconceptions about the LH partitioning method}

Wagg et al. seem to have fundamentally misunderstood the baseline that we proposed for the LH partitioning method. For example, they incorrectly claimed that our ``null expectation [...] should be that mixtures are overyielding to the degree to which species niches do not overlap and thus do not compete for the same resources'', and that our results were biased because we ``use[d] no correction for multiple niche overlap in mixtures of three or more species''. This suggests that they mistakenly believe that our baseline represents some type of application of the inclusion-exclusion principle. Additionally, if their statement about the need to correct for multiple niche overlap were true then it would mean that classical theory \citep[e.g.,][]{macarthur.levins_67} showing how the equilibrium abundances of (linear) multi-species community models could be calculated  by summing-up the pairwise effects would be wrong! In reality, it is Wagg et al.'s claim that is incorrect: our null expectation based on the sum of the pairwise effects does not require any correction for multiple niche overlap (\href{http://faraway.neu.edu/pillai_gouhier_pairwise.html}{see Appendix A for a numerical demonstration}).

Under the mistaken impression that our baseline did not correct for multiple niche overlap but LH's did, Wagg et al. went on to incorrectly claim that our ``null expectation is fundamentally different from the common null expectation used in LH, which corresponds to the first scenario in Fig. S1. Here the expected mixture yield is not the sum, but the average of all the component species' monoculture yields and thus increasing species richness does not change yield [\emph{sic}]''. This statement makes it seem as though the only difference between the LH null expectation and our baseline is that the former averages the monoculture yields whereas the latter sums them. However, this is only true when species do not interact in mixtures because only in that special case will pairwise effects be equivalent to the monoculture yields. Under the more likely scenario that species interact in mixtures, the total of the pairwise effects will not equal the total of the monoculture yields, so summing the pairwise effects represents an arithmetic operation applied to an altogether different quantity than that represented by the monoculture yields. Additionally, contrary to what the previous Wagg et al. quote suggests, averaging (rather than summing) the monoculture yields does not ensure that ``increasing species richness does not change yield [\emph{sic}]''. This statement is generally incorrect: it will only hold in the unlikely event that all species have identical monoculture yields.

Wagg et al. also appear to have confused the role of a null baseline (pairwise effects) with the reality it was designed to help measure (higher order effects). This led them to assume that our null based on pairwise mixtures is incomplete because it does not incorporate higher order competitive effects and thus cannot ``empirically [predict] species competitive outcomes at naturally occurring higher levels of diversity''. This is puzzlingly contrasted to the LH partitioning which is ``not theoretically limited by first having to parameterize all species' pairwise and higher order interactions in diverse plant mixtures''. However, contrary to what Wagg et al. claimed, our baseline only requires observations of ecosystem functioning in pairwise mixtures, not all mixtures of three or more species. Higher order effects can then be quantified by subtracting this baseline from the degree of ecosystem functioning observed in mixtures of three or more species.

Finally, Wagg et al. summarily dismissed our critique of the LH partitioning scheme by stating that it was never meant to ``provide direct evidence of \emph{any one particular} ecological mechanism underpinning BEF relationships as PG incorrectly assume'' (emphasis added). However, we never made such a claim. We simply demonstrated that under nonlinearity, the LH partitioning method will produce incorrect estimates of complementarity and selection, thus making it impossible for this statistical scheme to fulfill its intended purpose of quantifying the relative influence of \emph{broad categories} of ecological mechanisms on ecosystem functioning.

\section*{Dispelling misconceptions about diversity-interaction models}

As part of their criticism or our paper, Wagg et al. stated that the pairwise effects that we proposed as a null hypothesis ``[...] corresponds to what \citet{connolly.etal_13} use as an estimation of the [net] diversity effect - the sum of all the pairwise interaction effects and is actually just a special case of diversity-interaction models [...]''. However, this statement is incorrect because the pairwise effects in our framework are not equivalent to the pairwise interaction effects in diversity-interaction models. Indeed, although they have similar names, the pairwise effects in our framework and the pairwise interaction effects in diversity-interaction models represent distinct quantities that are used for different purposes. This becomes clear once one realizes that diversity-interaction models \citep{kirwan.etal_09,connolly.etal_13,brophy.etal_17} are simply repurposed General Linear Models that decompose variation in ecosystem functioning in terms of changes in initial overall abundance, species identity effects and species interaction effects. For instance, \citet{kirwan.etal_09} showed that ecosystem functioning $y$ could be decomposed as follows:

$$y = \alpha M + \sum_{i = 1}^{S}\beta_i P_i + \sum_{i<j}^{S} \delta_{ij} P_i P_j + \epsilon $$

\begin{figure*}[!htb]\centering
\includegraphics[width= 0.8 \linewidth]{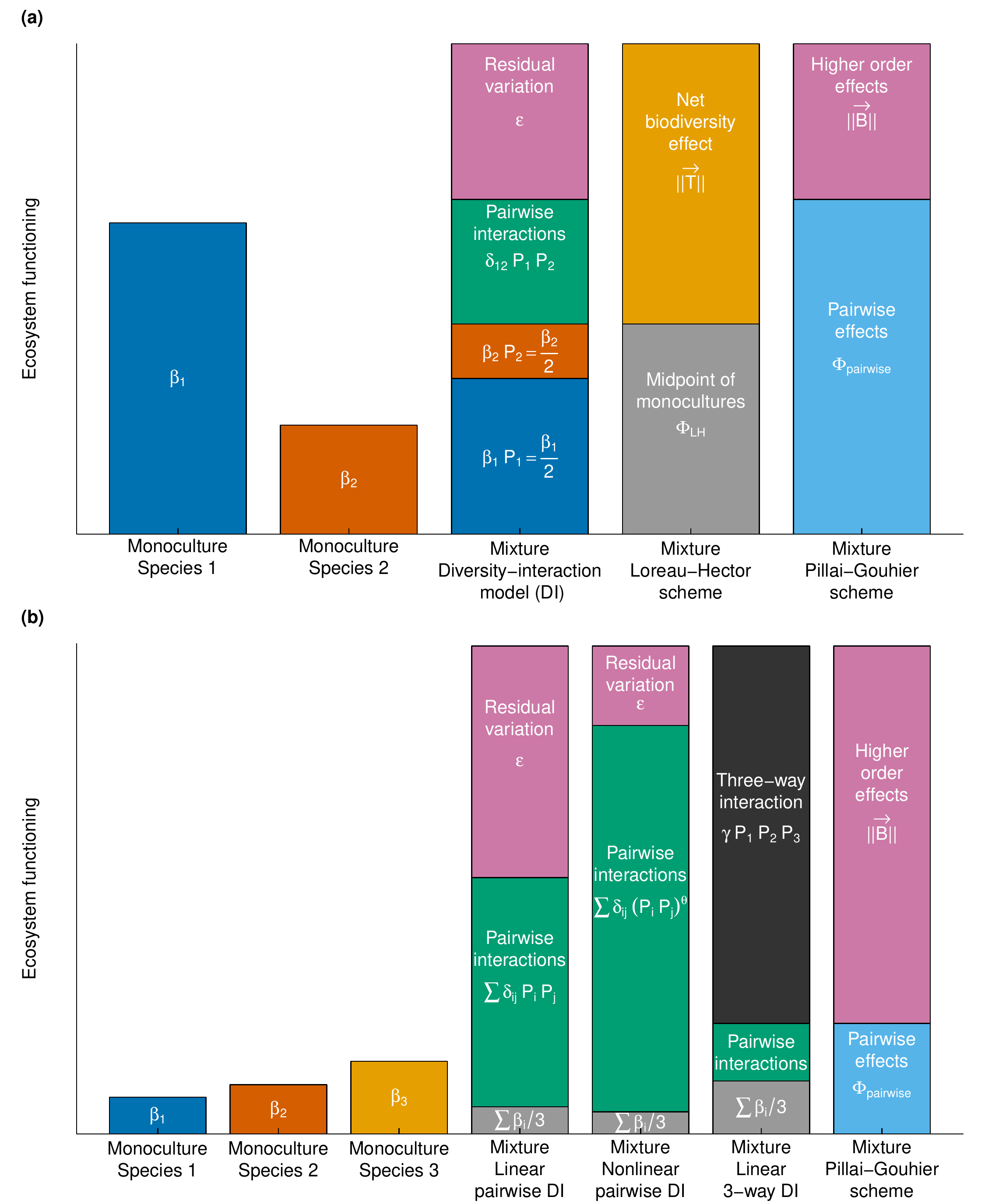}
\caption{\textbf{Dispelling misconceptions about diversity-interaction models and additive partitioning schemes}. (a) Diagram showing how ecosystem functioning is partitioned in a simulated community of two species using diversity-interaction (DI) models or additive partitioning with a baseline corresponding to the midpoint of the monocultures as suggested by \citet{loreau.hector_01} or the pairwise effects as suggested by \citet{pillai.gouhier_19}. (b) Misattribution of diversity effects to pairwise interactions in both linear and nonlinear diversity-interaction models in the presence of unaccounted for linear three-way interactions in simulated communities. The correct partitioning of ecosystem functioning is shown in the bar labeled ‘Linear 3-way DI' and was obtained by fitting a diversity-interaction model with species identity effects as well as linear pairwise and three-way interactions.}
\label{fig2}
\end{figure*}

\noindent where $M$ represents the total overall initial abundance and $P_i$ represents the proportional abundance of species $i$. Here, $\alpha$ represents the effects of changes in total initial abundance, $\sum_{i=1}^{S}\beta_i P_i$ represents identity effects (ID) for species 1 through $S$, $\sum_{i<j}^{S}\delta_{ij}P_i P_j$ represents the net diversity effect (DE) attributable to pairwise interactions $\delta_{ij}$ and $\epsilon$ represents the residual variation in ecosystem functioning.

If ecosystem functioning in mixtures results from the additive pairwise effects of species, the sum of the net diversity effect (DE) and the species identity effects (ID) estimated by fitting diversity-interaction models to monocultures and mixtures of 2 to $S$ species is equivalent to the pairwise effects obtained from 2-species mixtures in our framework (Fig. \ref{fig2}a). Diversity-interaction models thus essentially treat (a portion of) our null hypothesis as their alternate hypothesis by equating the net diversity effect to pairwise interactions. Hence, it is hard to understand how Wagg et al. can claim that our framework is ``[...] just a special case of diversity-interaction models'' and that ``[diversity-interaction models] are already much further developed than what PG proposed [...]'' when what we suggested was to delete pairwise effects in order to focus on the non-trivial component of ecosystem functioning that is due to higher order effects (Fig. \ref{fig2}a). Diversity-interaction models do the opposite: they quantify the net diversity effect via pairwise interactions (DE) after accounting for species identity effects (ID) and discard the rest as residual variation ($\epsilon$), whereas our approach discards the pairwise effects (ID + DE) in order to quantify higher order effects ($\epsilon$ in diversity-interaction models, Fig. \ref{fig2}a). If anything, by focusing exclusively on the trivial effect of pairwise interactions (DE), diversity-interaction models are `developing' in the wrong direction relative to the LH additive partitioning because the LH net biodiversity effect at least captures a portion of the non-trivial higher order effects (Fig. \ref{fig2}a). Overall, the only thing that diversity-interaction models and our framework have in common is the use of the term `pairwise effects', albeit at cross-purposes.

Based on the description above, one might think that our framework and diversity-interaction models \emph{could} be made equivalent by relabeling the sum of the species identity effects and the diversity effect (pairwise interactions) as `pairwise effects' and the residual variation as `higher order effects'. However, this will only work if the diversity-interaction model is specified correctly. For instance, if real communities are structured by linear pairwise and three-way species interactions, then fitting a diversity-interaction model based on pairwise interactions will yield incorrect estimates of both the species identity effects and the diversity effect (pairwise interactions), whereas our framework's estimates of both the pairwise effects and the higher order effects will remain accurate (Fig. \ref{fig2}b). This is because our pairwise effects are calculated using empirical observations obtained from pairwise mixtures only, whereas pairwise interactions in diversity-interaction models are estimated statistically from all the data (i.e., monocultures and mixtures of 2 to $S$ species). Hence, model misspecification can yield inaccurate inferences about the strength of the species identity effects and the net diversity effect.

Wagg et al. go on to suggest that ``[diversity-interaction models] already address the issues PG raised elsewhere [...]''. We assume that they are referring to the effects of nonlinearity in the LH additive partitioning method. If so, they are mistaken. \citet{connolly.etal_13} introduced so-called generalized diversity-interaction models to account for nonlinearity effects in mixtures by raising the product of the species abundances in the pairwise interactions to a power $\theta$. However, this type of generalized diversity-interaction modeling does nothing to address the fact that the nonlinear relationship between abundance and ecosystem functioning in monocultures inflates the LH net biodiversity effect and yields spurious estimates of selection and complementarity \citep{pillai.gouhier_19}. Not only does the inclusion of a nonlinearity parameter in generalized diversity-interaction models not address the issue that we described, in some ways, it actually exacerbates the problem. For instance, if the generalized diversity-interaction model is misspecified, the nonlinearity parameter will incorrectly attribute more of the variation in ecosystem functioning to nonlinear pairwise interactions when it should be attributed to linear higher order interactions (e.g., three-way interactions; Fig. \ref{fig2}b). Hence, generalized diversity-interaction models do not resolve the nonlinearity issue that we raised. If anything, the inclusion of a nonlinear parameter $\theta$ creates more attribution errors than it fixes under model uncertainty.

\section*{Confusing ecological principles with particular mechanisms}

Our critique regarding circularity focused on the ontological nature of species niches as deduced from basic ecological theory. We made no assumptions or claims about specific coexistence mechanisms or outcomes. When Wagg et al. argued about all the ways species coexistence is complicated and how its effects on ecosystem functioning are multifaceted, they were in essence making a category mistake: they confounded the general \emph{abstract} principles underlying coexistence with all of the \emph{particular} factors, processes and mechanisms that can play-out in in the real world to affect coexistence. They confused general tendencies in nature as delimited by scientific theories with all the specific and particular countervailing forces at work.

Even biological disciplines like ecology that have defied full axiomatization often still have basic principles that play a role similar to natural laws or tendencies; general propositions that essentially provide the organizing principles for defining concepts and expectations, and for determining methods in research -- even when the propositions do not always hold. The competitive exclusion principle and the notion of fitness stabilization/equalization serves as just such an organizing framework in ecology. The apparent refusal of the BEF research program to account for expectations grounded in these basic organizing principles is deeply problematic.

Indeed, by continuing to ignore basic ecological principles, the BEF research program runs the risk of becoming something analogous to a decades-long program that attempts to \emph{discern a general tendency of bodies to fall} by dropping different objects from, say, the top of buildings without accounting for gravity. As with BEF, circularity here would arise from the fact that a general propensity to fall is expected from our understanding of the \emph{built-in tendencies} of how the natural world is structured, and not because we assume \emph{a priori} that objects \emph{always} fall downward. One cannot respond to criticism that such a research program is trivial or circular by pointing out how the motions of released bodies are `multifaceted', and that the specific trajectories cannot be predicted due to the many distinct ways objects can move through space. The reality of bodies falling in different ways is not being denied; the multiple factors, mechanisms and processes involved in affecting the actual trajectories of bodies are not being erased. Nor does a call to account for the expected effects of gravity mean one is assuming that every dropped object will fall downward (after all, some objects may even fly or float away).

Here incorporating the \emph{ceteris paribus} assumptions of natural laws or tendencies is critical precisely because they allow us to explore and study this multifaceted behavior in order to understand how nature is structured and organized. For example, discounting expectations based on the effects of gravity when studying the motion of bodies could allow one to start disentangling the role of drag and friction in viscous fluids, discern the electric forces operating on charged particles in a field (as in the Millikan experiment), or understand the influence of air or other fluid currents on the motions or suspension of bodies in various mediums. Instead, for the last quarter century our proverbial research program has been content to endlessly point and gape with wonder at the simple fact that balls continue to roll \emph{downwards} along an incline.

When expectations based on basic principles are not accounted for, the use of statistical formalisms will likely serve to obscure these absurdities rather than enlighten. As a result, one risks mistaking trivialities for insights, as exemplified by an early classic paper that established how the ``motion of objects is extremely complex, subject to large numbers of influences. [...] Of the moving objects, the proportion moving down varied with size, temperature, wind velocity, slope of substrate if the object was on a substrate, time of day, and latitude. Among other things, this study concluded that ``On the whole, there is a slight tendency for objects to move down'' \citep{nabi_81}. These types of trivial conclusions will remain inescapable until BEF studies incorporate basic ecological principles into their underlying frameworks and expectations. Failing to do so will cause the field of BEF to devolve into the kind of degenerate research program that was decried decades ago by prominent ecologists.

\phantomsection
\section*{Acknowledgments}

We acknowledge support from the National Science Foundation (OCE-1458150, OCE-1635989, CCF‐1442728).

\phantomsection
\bibliographystyle{ecology}
\bibliography{pillai_gouhier_wagg_response}


\clearpage


\section*{Supplementary material}

# Appendix A: Demystifying pairwise effects

*Pradeep Pillai and Tarik Gouhier*

We created this R markdown document to show that there is no need to "correct for multiple niche overlap", as incorrectly suggested by Wagg et al. (2019; Ecology), when predicting the degree of ecosystem functioning in ecological communities. Here, like Loreau (2004; Oikos), we use a Generalized Lotka-Volterra (GLV) model to track the dynamics of species biomass, which represents our measure of ecosytem functioning. We begin by defining the number of species and the parameters of the GLV model:

```r
# Seed the random number generator for reproducibility
set.seed(20)
# Number of species
S <- 3
# Intrinsic rate of growth of each species
r <- rep(0.5, S)
# Carrying capacities of each species
K <- abs(rnorm(n = S, mean = 30, sd = 1))
# Interaction matrix
alpha <- matrix(abs(rnorm(n = S * S, mean = 1)), nrow = S, ncol = S)
# Set the intraspecific interaction coefficients to 1
alpha[row(alpha) == col(alpha)] <- 1
```

Next, we find the equilibrium biomass of each species based on the sum of the pairwise effects by inverting the interaction matrix and (matrix) multiplying by the carrying capacity vector as shown in Pillai & Gouhier (2019; Ecology) and countless previous studies:

```r
# Interior equilibrium biomass based on the sum of the pairwise effects
(N_equilibrium <- solve(alpha) %*% K)
```

```
##            [,1]
## [1,]   6.169808
## [2,]   9.191193
## [3,]  17.153702
```

Now, we write a function to numerically solve the GLV model in order to verify that the predictions based on the sum of the pairwise effects are accurate:

```r
suppressWarnings(library(deSolve))
# GLV model
solve_glv <- function(t, N, parms) {
  with(parms, {
    dN <- r * N * ((K - alpha %*% N) / K)
    return(list(dN))
  })
}
```

Finally, we use the deSolve package to numerically solve the GLV model in order to determine the equilibrium biomasses of the S species (solid curves) and compare them to the equilibrium biomasses predicted by the sum of the pairwise effects (dashed lines):

```r
# Same initial biomass for each species
N0 <- rep(0.1, S)
times <- 0:1000
parms <- list(r = r, alpha = alpha, K = K)
results <- ode(times = times,
               parms = parms,
               y = N0,
               func = solve_glv,
               method = "ode45")
# Plot the results
matplot(results[, 1], results[, 2:(S + 1)], t = "l", lty = 1,
        xlab = "Time", ylab = "Biomass",
        col = c("blue", "red", "darkgreen"),
        ylim = c(0, 25))
abline(h = N_equilibrium, lty = 3, col = c("blue", "red", "darkgreen"))
legend(x = "topright", lty = 1, col = c("blue", "red", "darkgreen"),
       legend = paste("Species", 1:3), bty = "n")
```

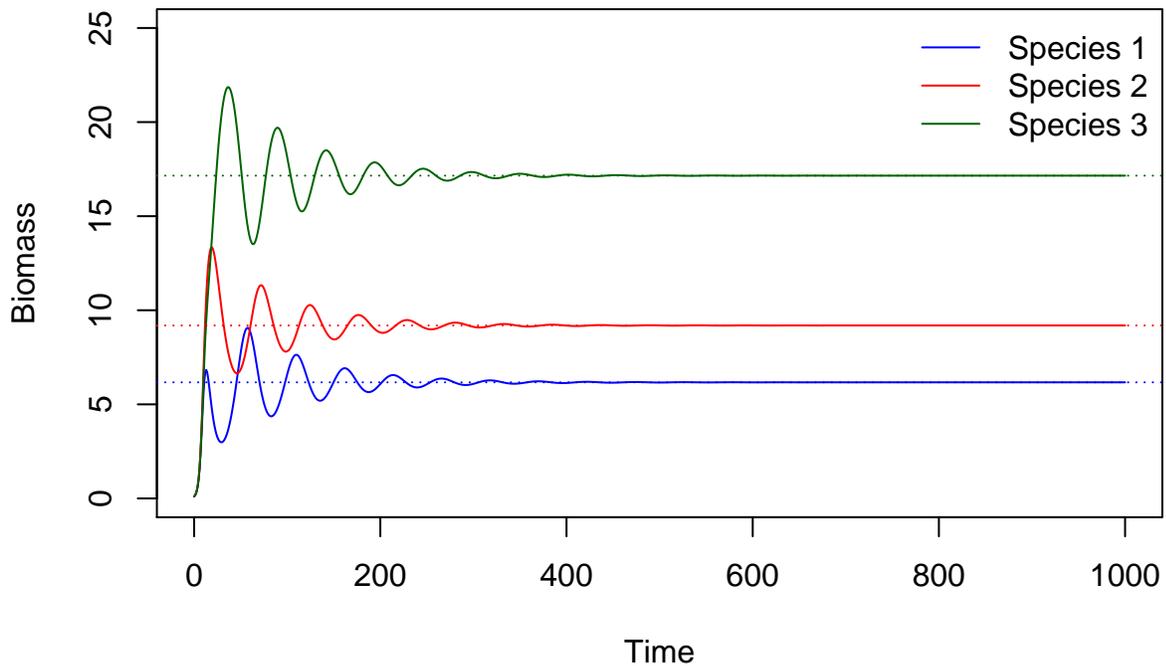

Unsurprisingly, the equilibrium biomasses predicted by the sum of the pairwise effects are equal to the ones generated by numerically solving the dynamics of the GLV model. Hence, contrary to what Wagg et al. (2019; Ecology) suggested, there is no need to "correct for multiple niche overlap" when using the sum of the pairwise effects to predict the degree of ecosystem functioning in multi-species communities.



\end{document}